\documentclass[aps,twocolumn,amssymb,floats,floatflt,prb,amsmath,superscriptaddress,showpacs]{revtex4}
\usepackage{graphicx}

\begin{document}

\title
  { Temperature and pressure-induced spin-state transitions in {LaCoO$_3$} } 

\author{Gy\"orgy Vank\'o}
\affiliation{European Synchrotron Radiation Facility, B.P. 220 F-38043 Grenoble Cedex 9, France }
\author{Jean-Pascal Rueff}
\affiliation{Laboratoire de Chimie Physique - Mati\`ere et Rayonnement (UMR CNRS 7614) Universit\'e Paris 6, 11 rue Pierre et Marie Curie, F-75231 PARIS CEDEX 05}
\author{Aleksi Mattila}
\affiliation{Division of X-ray Physics, Department of Physical Sciences, POB 64, FIN-00014, University of Helsinki, Finland}
\author{Zolt\'an~N\'emeth}
\affiliation{Department of Nuclear Chemistry, E\"otv\"os Lor\'and University, 
	 	P\'azm\'any s\'et\'any 1/A, H-1118 Budapest, Hungary}
\author{Abhay Shukla}
\affiliation{Institut de Min\'eralogie et de Physique des Milieux Condens\'ees, Universit\'e Pierre et Marie Curie, 4 place Jussieu, F-75252 Paris Cedex 05, France}

\begin{abstract}
We report the continuous variation of the spin moment of cobalt in LaCoO$_3$ across its temperature and pressure-induced spin transitions evidenced with $K\beta$ emission spectra. The first thermal transition is best described by a transition to an orbitally nondegenerate intermediate spin ($S$=1) state. In parallel, continuous redistribution of the $3d$ electrons is also indicated by partial fluorescence yield X-ray absorption spectra. At high pressure, our study confirms that the material becomes low spin between 40 and 70~kbar at room temperature.
\end{abstract}

\date{September 20, 2005}


\maketitle

\section{Introduction}

The perovskite-type cobalt oxide LaCoO$_3$ has attracted much attention since the 1950s due to its thermally induced spin-state and insulator-metal transitions. LaCoO$_3$ has a non-magnetic insulating ground state, with all the six $3d$ electrons of Co$^{3+}$ occupying the $t_{2g}$ level because the crystal field splitting exceeds the intraatomic exchange energy. Consequently, Hund's rule does not apply, and the cobalt is in low-spin (LS) state with $S$=0 for the $t_{2g}^6e_g^0$ configuration.
At higher temperatures two broad transitions have been observed in the magnetization measurements, occurring at around 90~K and 500~K. The classic interpretations invoked the appearance of the $t_{2g}^4e_{g}^{2}$ ($S$=2) high-spin (HS) state of Co$^{3+}$ at the first transition, and various ordered superstructures of the HS and LS cobalt atoms at higher temperatures.\cite{goodenough1958,raccah1967} Inconsistencies started to arise, however, when neutron diffraction studies failed to find any superstructures,\cite{thornton1986} and more conflicting discrepancies were brought up by experimental studies of the electronic structure. Namely, it was found that neither Co $2p$ and O $1s$ X-ray absorption spectroscopy (XAS) nor photoemission spectra show pronounced variations below 300~K; they reflect a gradual transition only above 500~K from the LS state to a mixed spin state with a main contribution from the HS state.\cite{abbate1993} The failed observation of the 90-K transition and the conflicting description of the temperature behavior stimulated significant further efforts to determine the magnetic and spin states of LaCoO$_3$. Shortly after these experiments, the 90-K transition was evidenced by a spin-polarized neutron-scattering study.\cite{asai1994} Not much later, a completely different description was proposed that included the formation of a $t_{2g}^5e_g^1$ ($S$=1) intermediate-spin (IS) state at the first transition.\cite{potze1995} Theoretical works completed this LS/IS scenario with proposing an orbital ordering for the IS state, which melts at the second transition.\cite{korotin1996,ravindran2002} This has become the prevailing scenario, and most of the recent experimental observations are interpreted as supporting the IS state.\cite{noguchi2002,zobel2002,maris2003} Although including this state in the explanations seems to explain the observed phenomena, some authors still invoke the HS state at higher temperatures.\cite{asai1998,louca1999,radaelli2002,rao2004,kyomen2005} Another sign of the incompleteness of a simple LS/IS scenario is that the description of the high temperature susceptibility data usually needs a treatment different from the low temperature ones, e.g. invoking change in the orbital degeneracy at the second transition.\cite{zobel2002} Thus, the interpretations of the two transitions are still conflicting; the precise spin state as a function of temperature, and the variations of the electronic structure associated with the transitions are not yet fully understood, and require further studies.

Pressure can also cause spin-state transitions in transition metal compounds,\cite{pasternak1997} as the crystal field splitting rapidly increases with the shrinking of the bond lengths ($10~Dq\propto~r^{-5}$). Since the transitions of LaCoO$_3$ are accompanied by anomalous volume expansions, it is apparent that relatively low pressure can have considerable effect on the spin state. The effect of pressure to the 90-K transition was studied up to 18 kbar by Asai \textit{et al.}, who found that the energy gap between the LS and the higher spin state increases with pressure.\cite{asai1997} Recently, Vogt \textit{et al.} performed an X-ray powder diffraction study to examine the pressure-induced changes. The results were interpreted as a continuous transition from IS to LS state, which is completed at around 40~kbar.\cite{vogt2003} Chemical pressure, introduced by partial substitution of La$^{3+}$ with the smaller Eu$^{3+}$, leads to similar stabilization of the LS state.\cite{baier2005}

We present here high-resolution X-ray spectroscopy study of the spin state and the electronic structure of cobalt in LaCoO$_3$ as a function of temperature and pressure. The applied techniques allowed us to follow the occupancy and the local magnetic moment of the cobalt $3d$ orbitals. The applicability of the X-ray absorption near-edge structure (XANES) spectroscopy to study the electronic structure is well known, as it gives access to the unoccupied density of states. In the case of the spin transition, it is expected that the pre-edge of the main $1s\rightarrow4p$ absorption line, which has a relevant contribution from quadrupolar $1s\rightarrow3d$ transitions, follows sensitively the variations in the populations of the $t_{2g}$ and $e_g$ levels.\cite{vanko2002} These orbital populations are expected to undergo relevant changes, as the spin transition in a localized picture is best described as redistribution of the electrons between the $t_{2g}$ and $e_g$ orbitals, as it was described above. The applied partial fluorescence yield (PFY) detection technique,\cite{hamalainen1991} even if it is not a true measure of the absorption cross section,\cite{carra1995} allows better energy resolution than conventional XANES,\cite{glatzel2005} and can be used to monitor more sensitively the population of the different states.\cite{dallera2002,glatzel2005} X-ray emission spectroscopy (XES), in fact, the $K\beta$ emission line can be utilized to probe the local magnetic moment on the $3d$ orbitals, i.e., the localized spin moment.\cite{rueff1999prl,vanko2002,badro2004} The origin of the sensitivity to the spin state is the strong exchange interaction between the $3p$ hole and the $3d$ orbitals in the final state of the emission process; the two different possible orientations of the spin of the $3p$ hole with respect to the resultant local moment of the unpaired $3d$ electrons leads to a splitting.\cite{tsutsumi1976} The arising spectra are not simple due to the numerous multiplet terms involved; nevertheless, a low-energy satellite ($K\beta'$) to the main emission line ($K\beta_{1,3}$) is observed, whose intensity is proportional to the spin localized on the $3d$ orbitals.\cite{tsutsumi1976} Besides this phenomenological description, a full theoretical treatment based on crystal-field or charge-transfer multiplet theory can satisfactorily reproduce the spectral variations with the spin state.\cite{peng1994jacs,wang1997}

\section{Experiment}

LaCoO$_3$ was synthesized as described in Ref.~\onlinecite{zobel2002}. 
For reference samples, high purity LiCoO$_2$ and CoF$_3$ were purchased from Sigma-Aldrich.
The experiment was carried out at the ID16 beamline of the European Synchrotron Radiation Facility. 
The undulator radiation was monochromatized with a cryogenically cooled Si(111) double-crystal monochromator, and focused to a spot size of 150$\times$100 $\mu$m$^2$ with a Rh-coated toroidal mirror. 
Arranged on a Rowland-circle spectrometer of 1 m diameter, a spherically bent Si(620) crystal and a Peltier-cooled Si diode were applied to analyze and detect the emitted radiation, respectively. The $K\beta$ spectra were collected by scanning the Bragg angle of the analyzer around 70.69$^\circ$, which corresponds to the maximum of the Co $K\beta$ emission line; the detector was also translated and rotated accordingly. The overall resolution of the spectrometer was 1.5 eV.
Fluorescence yield XANES spectra were taken with a photon-counting R\"ontec detector. Partial fluorescence yield XANES spectra were taken on the maximum of the $K\beta$-emission line by scanning the incident energy through the absorption edge (and keeping the analyzed energy constant).\cite{hamalainen1991} For the temperature dependent studies, a scattering angle of 90$^{\circ}$ was chosen.
The low-temperature experiment was carried out in a standard closed-cycle helium cryostat. For the high-temperature studies, the sample was mounted on a Ta/Mo support heated by a ceramic heater device, in a vacuum cell.

A Mao-Bell diamond anvil cell with a Re gasket was used for the high-pressure experiment. The sample was prepared by grinding a piece of a LaCoO$_3$ crystal, and a methanol-ethanol (4:1) pressure medium was added to the powder sample to ensure hydrostatic pressure. The pressure was determined by the ruby fluorescence method. Both the incident and the scattered radiation passed through the diamond anvils; the scattering angle was 30$^{\circ}$. In order to minimize loss of the emitted radiation, a diamond of smaller height was used on the downstream part of the cell: the diamonds had heights of 2.5 mm and 1.5 mm, with culet sizes of 0.7 mm and 0.6 mm, respectively. The energy of the exciting beam was set to 13 keV, a compromise between the absorption loss in diamond and scattering cross section.

All spectra were normalized to the beam monitor; the XES spectra were further normalized to the spectral area, and the spectra were aligned to have the main lines at the same energy.

\section{Results}

\begin{figure}  
\includegraphics[width=8.5cm]{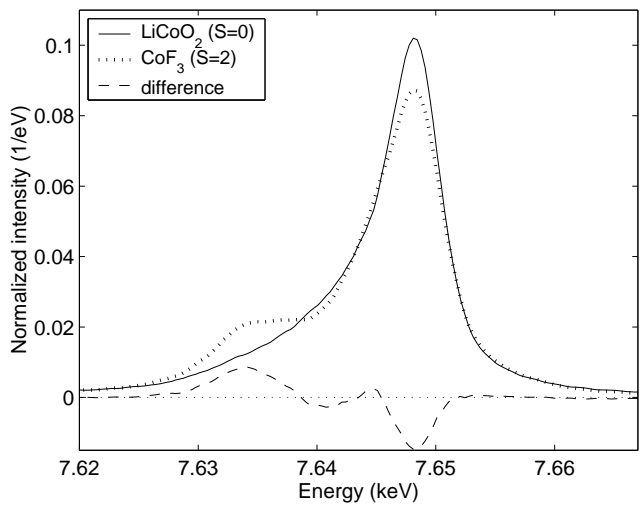}
\caption{\label{figco3ref} $K\beta$ spectra of LiCoO$_2$ and CoF$_3$, a LS and a HS Co$^{3+}$ compound, and their difference.}
\end{figure}

Figure \ref{figco3ref} shows the XES spectra of our HS (CoF$_3$) and a LS (LiCoO$_2$) reference compounds. As it was expected from previous observations,\cite{rueff1999prl,vanko2002} and theoretical calculations,\cite{wang1997} the largest spectral variations are seen on the intensity of the main $K\beta_{1,3}$ line and on its $K\beta'$ low-energy satellite (see the difference spectrum in Fig.\ \ref{figco3ref}). 

\begin{figure} 
\includegraphics[width=8.5cm]{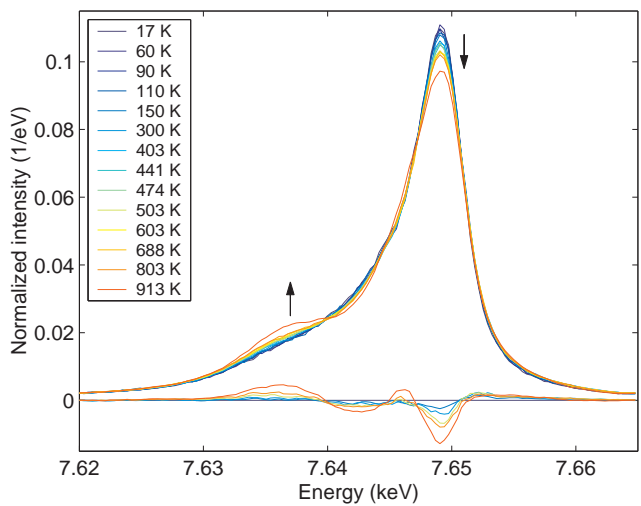}
\caption{\label{figkb} (color online) Evolution of the LaCoO$_3$ $K\beta$ spectra with temperature, arrows indicate the main line shape variations with increasing temperature. At the bottom, differences from the lowest temperature spectrum are shown for 17 K, 150 K, 300 K, 503~K, 803~K, and 913 K.}
\end{figure}

Similar variations, though to a lesser extent, were observed in the line shapes of the spectra of LaCoO$_3$ (Fig.\ \ref{figkb}), as a function of temperature. At the lowest temperatures the spectra are identical to that of LiCoO$_2$. As temperature increases, the increase of the satellite intensity and the decrease of the main line intensity is observed. These variations are continuous with temperature, which reflects a continuous change of the average spin of the cobalt. The 913-K spectrum exhibits a sudden change. While below this temperature the spectral variations were reversible, the last spectrum remained the same when the sample was cooled down after collecting data at 913 K. This indicates that a chemical transformation took place when reaching the highest temperature of our study; this will be discussed later in more detail.

\begin{figure} 
\includegraphics[width=8.5cm]{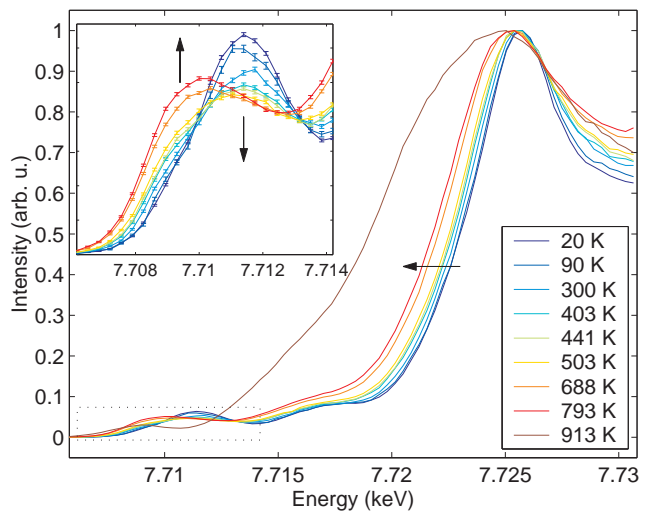}
\caption{\label{figpfy} (color online) PFY XANES of the LaCoO$_3$ on the max of $K\beta$, with the pre-edge region in the inset. Arrows indicate the spectral variations at increasing temperature.}
\end{figure}

The PFY XANES, displayed on Fig.~\ref{figpfy} shows also continuous variation with temperature. 
Below the $1s\rightarrow4p$ main absorption line at 7726 eV, several low intensity features appear. Following earlier interpretations,\cite{kim1997,toulemonde2001,haas2004} the broad feature at around 7716 eV has been assigned to a ligand-to-metal charge transfer (LMCT) shake-down process, which results in a final state of $1s^13d^74p^1\underline{L}$, instead of the $1s^13d^64p^1$ of the main transition, ($\underline{L}$ denotes a hole on the ligand). At even lower energies, the pre-edge region reflects transitions to the $t_{2g}$ and $e_g$ levels, partly through direct quadrupolar transitions, but also through to dipolar transitions to d-p hybridized orbitals. The latter will affect the pre-edge intensity but not the spectral shape, which still shows characteristic features of the $3d$ crystal-field splitting even in the dipolar component.\cite{arrio2000} At higher temperatures, the edge progressively shifts while the pre-edge region, as it will be discussed later in more detail, reveals a continuous redistribution of electrons on the $t_{2g}$ and $e_g$ levels through intensity changes, such as broadening and the shift of their spectral weight to lower energies. At 913~K, similarly to the XES, a drastic change was observed in the spectrum: the absorption edge was shifted by several eVs to lower energies, and the pre-edge was also shifted and reduced. The large edge shift indicates a change in the valence state of cobalt, which was confirmed by chemical analysis: the sample was transformed to LaCoO$_{2.5}$, which reveals that a Co$^{3+}\rightarrow$ Co$^{2+}$ transformation and simultaneous oxygen loss took place. This is not entirely surprising, as Radaelli and Cheong have found a third lattice anomaly (besides the two at the magnetic transitions) around 800~K, which was attributed to formation of oxygen vacancies.\cite{radaelli2002} Moreover, such reduction is known to happen to LaCoO$_{3}$ at even lower temperatures in the presence of a reducing agent.\cite{gilbu1995}

\begin{figure} 
\includegraphics[width=8.5cm]{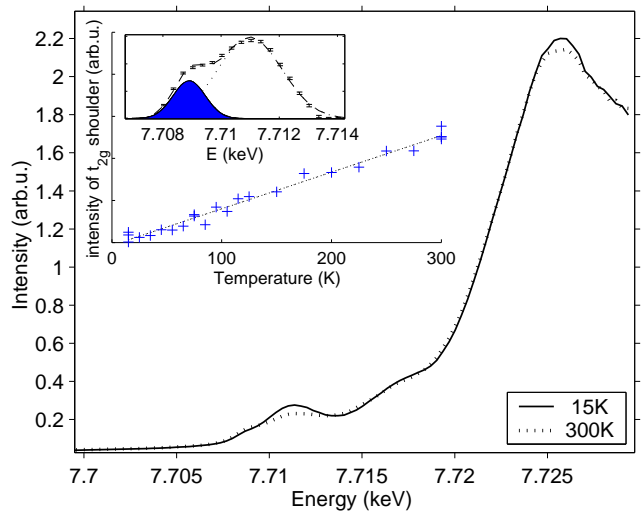}
\caption{\label{figCoK} TFY XANES at 15 K and 300K. The intensity variation of the low-energy ($t_{2g}$) shoulder in the pre-edge of the Co K-edge is shown in the inset. Above this plot, a spectrum taken at T=300K illustrates (after background subtraction and normalization) the fitting with which the plotted values are obtained.}
\end{figure}	

As only a few spectra were taken at low temperatures, rapid total fluorescence yield XANES scans were measured between 15 K and 300~K to verify that the spectral changes (hence the redistribution of the $3d$ electrons) are indeed continuous in this region. The spectra are shown in Fig.~\ref{figCoK}; the lowest and highest-temperature spectra are similar to those of Toulemonde (taken at similar temperatures),\cite{toulemonde2001} in between a continuous change is seen as displayed in the inset of the figure. 
The pre-edge was interpreted as a result of two contributions:\cite{toulemonde2001,haas2004} a low-energy feature of small, but increasing intensity was attributed to transitions to the $t_{2g}$ orbitals, while the diminishing larger feature at higher energy was attributed to $e_g$ orbitals. The relative change of their spectral weight is therefore a sign of the $t_{2g}\rightarrow e_g$ electron transfer. Although this is a simplified description, the nature of the changes of the electronic structure indeed originate from the redistribution of the $3d$ electrons; therefore, for monitoring purposes this description should suffice. As we see in the inset of Fig.\ \ref{figCoK}, which presents simple two-Gaussian fits to the pre-edges, the variation of the features of the pre-edge are indeed continuous. To derive a quantitative description of the redistribution of the $3d$ electrons, a detailed resonant X-ray emission study would be desirable.

\begin{figure}  
\includegraphics[width=8.5cm]{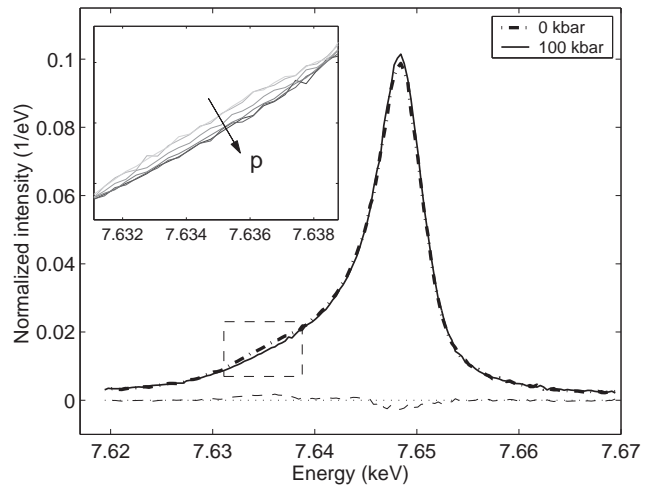}
\caption{\label{fighpkb} Evolution of the LaCoO$_3$ $K\beta$ spectra with pressure. The main figure shows the extremes and their difference, while the inset displays the $K\beta'$ satellite region for all the measured pressure points, which include 0, 7, 22, 42, 70, and 100 kbar.}
\end{figure}

As for the pressure dependence, a gradual variation of the $K\beta$ line shape was found up to 70~kbar. As it is seen in Fig.\ \ref{fighpkb}, the relative intensity of the satellite were reduced while that of the main line were increased, which is typical for a transition toward the LS state. Above 70~kbar no further changes were observed. In this case we could not monitor the redistribution of the $3d$ electrons directly with XANES due to the strong absorption of the diamonds at the Co K edge. However, the nature and the extent of the changes in the XE spectra between ambient pressure and 70 kbar are the same as those which are observed when the temperature is decreased at ambient pressure from 300 K to 17 K; this similarity will be more obvious in section \ref{IVB}, where this comparison can be made in a more quantitative manner. At 70~kbar and above the spectra appear the same as the lowest temperature one of the temperature dependent series. This suggest that at room temperature a pressure-induced transition to the LS state is essentially completed between 40~kbar and 70~kbar, which is in agreement with the findings of Vogt \textit{et al}.\cite{vogt2003}
   
\section{Discussion}

\subsection{Model calculations}

In order to facilitate the understanding of the variation of the
experimental data and to help to relate it to the spin state, the Co $K\beta$
emission and the pre-edge at the K-edge of Co in LaCoO$_3$ were modeled using a
charge transfer multiplet
approach \cite{deGroot1994,deGroot2001}. The XANES  spectra were also
calculated using an \textit{ab initio} multiple scattering code.

The charge transfer multiplet model used a $3d^6+3d^7\underline{L}+3d^8\underline{L}^2$ 
ground state basis set, where $\underline{L}$ denotes a hole in the ligand
valence band. The Slater integrals were obtained by the Hartree-Fock method 
\cite{cowan1981} and scaled by 0.8 to take into account intra-atomic configuration
interaction effects. For LS couplings atomic values were used. The crystal field 
perturbation in $O_{h}$ symmetry was considered using the approach developed by Butler\cite{Butler1981},  
and charge-transfer effects were taken into account by using the code by Thole and Ogasawara\cite{FDGweb}. The $1s$ lifetime was set to 1.3 eV and the final state lifetime broadening FWHM was
approximated as $0.2~\times~E_{f}$, where $E_f$ is the final state energy \cite{taguchi1997jpsj}. 

Although the IS state cannot be obtained as a true ground state in the multiplet approach, it was assumed that using the triplet $^3$T$_1$ term of $3d^6$ can provide a reasonable approximation for the spectra of
this state. The model parameters for the different spin states of LaCoO$_3$, as well
as for a high-spin and a low-spin reference compound of $3d^6$
configuration, CoF$_3$ and LiCoO$_3$, are presented in
Table~\ref{tabcalc}. The parameter values were obtained from the literature,\cite{saitoh1997,abbate1994,chainani1992,matsushita2002,mizokawa1996}
and they were modified to some extent when necessary to achieve better agreement with the
experimental data. The calculation with the LS ground state was compared 
to the experimental spectrum of LaCoO$_3$ measured at 17 K and to that of LiCoO$_2$. The calculation on the intermediate spin ground state was tentatively compared with the experimental spectrum measured at 403 K, and  CoF$_3$ was used to represent also LaCoO$_3$ in the high-spin state. In the table $\Delta$ stands for the charge transfer energy, defined as the energy difference between the centers of gravity of $d^{n}$
and $d^{n+1}\underline{L}$ configurations and $U_{dd}$ is the Hubbard $U$
given by $E(d^{n+1})+E(d^{n-1})-2E(d^{n})$. The core hole potential $U_{dc}$ 
is set to the same value for both $1\underline{s}$ and $3\underline{p}$ holes.
$V_{e_g}$ is the hybridization strength for the $e_g$ orbitals in the ground state; the 
hybridization strength for $t_{2g}$ states was set to half of the $V_{e_g}$ value.
For states with a core hole the $V_{e_g}$ hybridization strength was reduced by $0.4$ eV.

\begin{table}
\caption{\label{tabcalc} parameters of the multiplet calculations. For
description of parameters see the text. }
\begin{ruledtabular}
\begin{tabular}{llllllll}
Spin & Materials  & $\Delta$ & $U_{dd}$ & $U_{dc}$ & $Ve_g$ & 10$dq$ & ~
\\ \hline

$S$=0 & LiCoO$_2$, LaCoO$_3$ LS    & 5 & 6.5 & 7 & 1.8 & 2.0 \\
$S$=1 & LaCoO$_3$ IS & 4 & 6.5 & 7 & 2.0 & 1.8 \\
$S$=2 & CoF$_3$, LaCoO$_3$ HS 	    & 1 & 5.0 & 7 & 2.2 & 0.8 \\
\end{tabular}
\end{ruledtabular}
\end{table}

\begin{figure}  
\includegraphics[width=8.5cm]{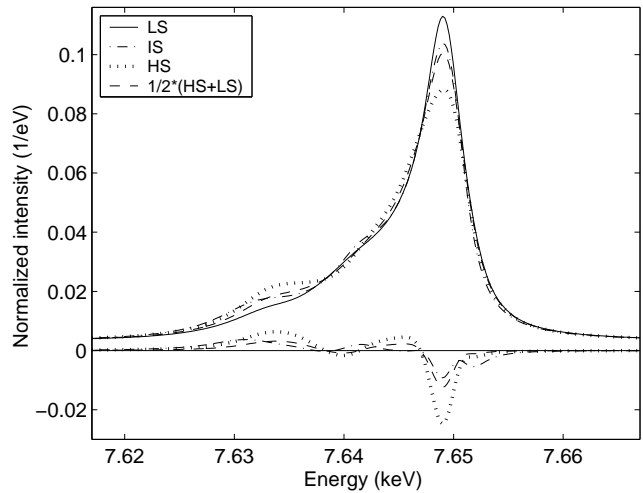}
\caption{\label{figcalckb} Calculated $K\beta$ spectral shapes of LaCoO$_3$
for the different possible spin states; the mean of the HS and LS spectra is also shown for comparison with that to the IS. The differences from the LS spectrum are also plotted in the bottom of the figure.}
\end{figure}

The simulated $K\beta$ spectra for the different spin states of LaCoO$_3$ are shown on Fig.~\ref{figcalckb}. The spectra are shifted so that the main lines match the nominal Co $K\beta$ energy. In agreement with the experimental data and with previous observations\cite{rueff1999prl,vanko2002} and calculations,\cite{wang1997} the intensity of the $K\beta'$ low-energy satellite of the main $K\beta_{1,3}$ line is increasing with the increasing cobalt spin moment. 
However, there are no apparent spectral variations that would provide an easy way to verify the presence of the IS state. According to Tsutsumi's proposal, the position of the satellite with respect to the main peak should move with the spin state as $\Delta E=J(2S+1)$ (where $J$ is the $3p-3d$ exchange integral),\cite{tsutsumi1976} thus for the IS a substantial upward shift (almost 6 eV) of the $K\beta'$ feature would be expected. However, as observed also at other spin transitions,\cite{vanko2002} this is not the case: no new features show up, and the calculated IS spectrum barely differs from the mean of those of the HS and LS (\textit{cf.} Fig.\ \ref{figcalckb}). This is also in accord with the temperature dependent experimental data on LaCoO$_3$ provided that the first transition is indeed an LS$\rightarrow$IS one, since no relevant energy shifts of the satellite were observed. 
In general, the variations of the experimental spectra (when available) are well accounted for by the calculations. The calculated line shapes suffer from only a slight overestimated height of the a $K\beta'$ satellite, which originates from the approximations of the lifetime broadening model and the overestimation
of transition matrix element strengths in the calculation. However, due to the discussed problems, the satisfactory agreement is not sufficient to extract quantitative information about the spin state or discriminate between LS/HS or LS/IS scenarii in LaCoO$_3$. We will show that careful line shape analysis can overcome this shortcoming (\textit{cf.} section \ref{IVB}).

\begin{figure} 
\includegraphics[width=8.5cm]{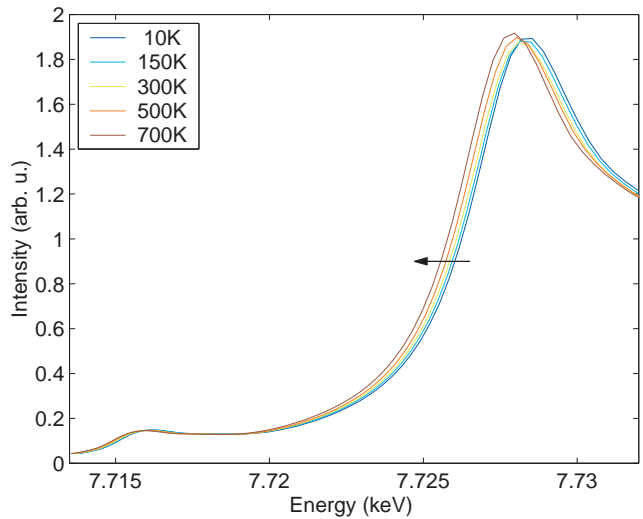}
\caption{\label{figcalcfeff} (color online) Co K edge of LaCoO$_3$ calculated with
\textsc{feff8.2} using structural data from literature.\cite{radaelli2002} An arrow indicates the variations at increasing temperature.}
\end{figure}

For the description of the Co XANES spectra, an \textit{ab initio} multiple scattering approach was used to study the spectral variations with the temperature for a few selected values between 10~K and 700~K. The temperature dependent structural data was taken from the literature\cite{radaelli2002}. The
program utilized was \textsc{feff}8.2,\cite{ankudinov1998,ankudinov2002} a
real-space multiple scattering code. For the calculations the Hedin-Lunqvist
self-energy and a cluster of 134 atoms was used. Dipolar and quadrupolar transitions 
were both considered to get a
reasonable estimation for the pre-edge. As seen from the simulated
spectra shown on Fig.\ \ref{figcalcfeff}, the calculations reasonably
describe the edge shifts, which accompany the structural changes with
temperature. However, as expected for a one-electron approach, it fails to reproduce the LMCT peak, and does not show relevant variation on the pre-edge. Therefore, for a better
description of the pre-edge region a different theoretical approach,
the charge transfer multiplet model was invoked. 
The quadrupole contributions were calculated using the multiplet code with the same parameters as
presented in Table \ref{tabcalc}. The dipolar part was modeled using a phenomenological
density of $p$ symmetry final states similar to the model used by Caliebe \textit{et al.}
\cite{caliebe1998} in calculating the pre-edge of Fe $K$-absorption edge of Fe$_2$O$_3$.
The charge transfer multiplet calculation predicts a single peak in the pre-edge for the LS
state, which originates from transitions into $e_g$ levels. 
When higher spin states are populated, the lower-lying $t_{2g}$ levels also become accessible at the absorption
process due to their partial depopulation. The appearing features are expected
to cause a 'widening' of the pre-edge, especially in the direction toward
lower energies. This is in agreement with the experimental data presented
in Figs. \ref{figpfy} and \ref{figCoK}. The determination of the spin
transition on the basis of this spectra is again challenged by the
relatively small differences of the IS and HS line shapes.

\begin{figure} 
\includegraphics[width=8.5cm]{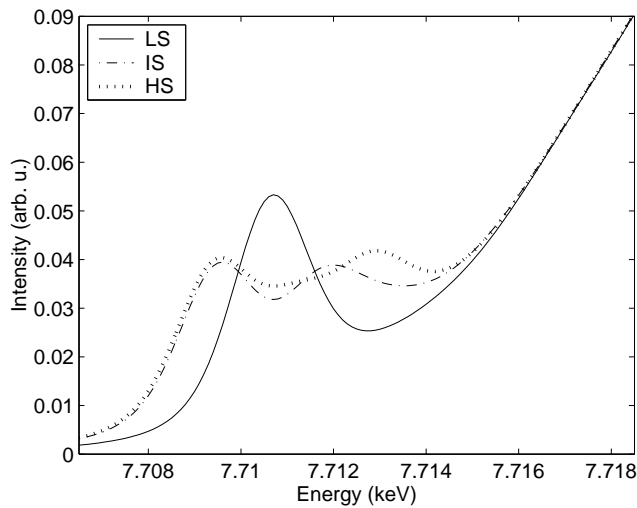}
\caption{\label{figcalcpreedge} Results of the multiplet calculations on the pre-edge of the Co K-edge of LaCoO$_3$ for the different spin states}
\end{figure}

\subsection{Line shape analysis}
\label{IVB}

Despite the conspicuous effects of the localized spin on the $K\beta$ spectra, and the evident origin of the satellite, the task of determining the spin from the line shape is not obvious. As multiplet theory shows, the spectra are made up by contributions of numerous terms, and the arising lines can even have different widths due to the mentioned different term-dependent lifetimes. Therefore, fitting the spectra with theory would lead to many parameters whose number might be close to that of the data points. 
For a phenomenological description, the $K\beta$ spectra can usually be fitted reasonably with three Voigt functions.\cite{gamblin2001} This might even serve as a useful approach for examining trends over a large set of compounds, for example; nevertheless, it has little use in quantifying the spin state for a transition which is not completely known.\\
A more effective approach is to use the combination of proper reference spectra of known spin states of transition metals with similar local structure to model the spectral shape, or eventually to fit it to the data. In such cases the spectra can be reasonably reproduced, and the extent of the transition can be procured, as it was demonstrated in Ref. \onlinecite{badro2004}. However, this requires that the reference spectra is taken under similar conditions, i.e., similar resolution, source size, background, and it is seldom that these are all satisfied. Moreover, it might not always be easy to find appropriate reference materials, e.g., for the present study no IS Co$^{3+}$ compound was available, and using the CoF$_3$ as HS reference can also give less satisfactory results for a CoO$_6$ system due to the differences in the bonding, let alone the uncertainties faced when interpolating the line shapes for the not entirely known IS state.
So far the most successful approach is to follow the variation of the spin state with this technique was to plot the integrals of the absolute values of the difference spectra (\textit{hereafter} IAD), and then scale them to values deduced from some known reference materials.\cite{rueff2001} The IAD value for spectra SP$_i$ with respect to a reference spectra SP$_{ref}$ is obtained as numerical approximation to Eq.~\ref{eqIAD}:

\begin{equation}
\textrm{IAD}_i = {\int |\ \textrm{SP}_i-\textrm{SP}_{ref}\ |\ dE},
\label{eqIAD}
\end{equation}

(Note that this integral would give 0 without taking the absolute values due to the normalization of the spectra.) This approach deserves further consideration as it has many advantages. It is model independent, it can be performed without invoking results of theoretical calculations, it relies only on proper reference spectra \textit{or} IAD values obtained for a similar transition, and the raw data needs only a simple preparation before the analysis can be used, so it is a promising tool for the rapid determination of the spin state from the XES spectra. On the other hand, it requires that the whole set of experimental data are acquired under similar conditions.

The area differences were so far used for the determination of the spin state in a pragmatic manner.\cite{rueff2001} Nevertheless, it can be easily demonstrated that in case of a transition between two states, this approach, although it might look arbitrary at first, is a true probe of the line shape variation, which is parallel to the variation of the spin.\footnote{For this, it should be confirmed that the IADs and the line shape changes are correlated: it will be shown that the IAD values of linear combinations of any two normalized functions are proportional to the IAD value of the two functions, and the proportionality factor is the weight in the linear combination. Let us consider the functions $f(x)$ and $g(x)$, which are normalized to give unit area at integration. Consequently, the integral of their difference $(f-g)$ is zero, as integration is a linear operation. The corresponding IAD value (let us note it $\alpha$), however, is different from zero if $f\neq g$; thus $\alpha=\int{|f(x)-g(x)| dx}$. A spectrum of an intermediate state can be written as a superposition of $f$ and $g$: $ SP=c_1 f + c_2 g$, where $c_1 + c_2 = 1$. Its difference from the reference $g$ is $SP-g=c_1 (f - g)$. The integral of the absolute value of the latter is IAD = $c_1\int{|f(x)-g(x)|dx}=c_1 \alpha$; this proportionality to the full area difference is what makes the IAD value a good measure of the extent of the transformation.}

\begin{figure} 
\includegraphics[width=8.5cm]{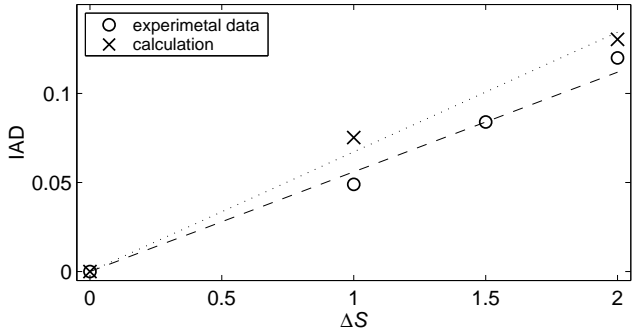}
\caption{\label{figIADvsSpin} IAD values for the calculated and the measured reference spectra.}
\end{figure}

In the present case, however, three spin states might be involved in the transitions. In order to follow the spin state through the line shape variations, we have to examine, whether and how the IAD correlates with the spin. Unfortunately, we have no means of expressing this in form of analytical functions, we can only rely on values obtained from the simulated and the experimental line shapes. In Fig.~\ref{figIADvsSpin} the experimental IAD values for the $\Delta S$=0, 1, 3/2, and 2, as well as the IADs derived from the calculated spectra of  Fig.\ \ref{figcalckb} for the $S$=0, 1, and 2 states are presented. 
The IAD obtained for the CoF$_3$ HS reference compound (compared to the LiCoO$_2$ LS reference) is 0.12 for the $\Delta S$=2; LaCoO$_{2.5}$, which is known to have S=3/2, gives an IAD of 0.084. Finally, for a molecular spin-transition system of Co$^{2+}$ with a $\Delta S$=1 transition an IAD of 0.049 was obtained.\cite{vanko2002} Apparently both IAD sets vary linearly with the spin, thus on the basis of the existing data it seems to be at least a promising approximation to determine the spin state. The linearity possibly arises from the the similarity of the environments of Co, as the ligands provide a somewhat distorted octahedral surroundings in all cases, and it is well known in multiplet theory that the local symmetry is more important to determine the spectral shape than the nature of ligands or the valence state.\cite{wang1997} The larger values of the calculated ones can be explained by the overestimation of the $K\beta'$ intensity mentioned above; rescaling this data to the experimental one using the fraction obtained from the proportion of the slopes of the fitted lines, suggests that the calculation overestimates the line shape variation by 17\%.

\begin{figure} 
\includegraphics[width=8.5cm]{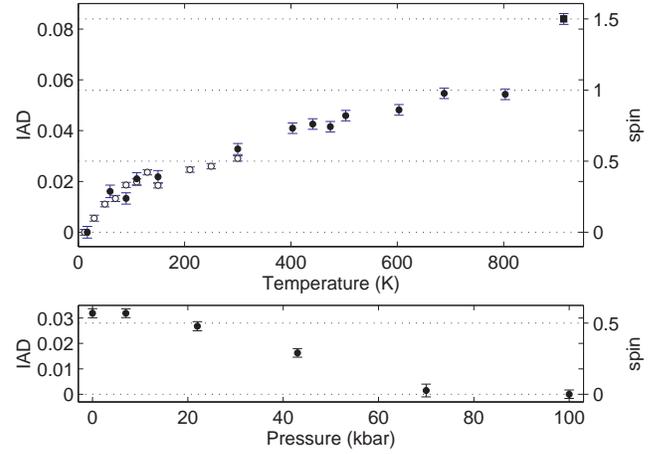}
\caption{\label{figarea} Integrals of the absolute values of the area differences of $K\beta$ spectra. Top: temperature dependence; the open and filled symbols represent data from two different samples. Bottom: pressure dependence at room temperature. The dotted lines represent one half increments in the spin value.}
\end{figure}

We followed this approach to analyze the data; the IAD values obtained with Eq.~\ref{eqIAD} are presented in Fig.\ \ref{figarea}. In order to convert these values to spin, the linear dependence derived from Fig.~\ref{figIADvsSpin} were used to rescale the data to determine the average local spin of cobalt as a function of temperature or pressure. The obtained spin values are presented on the right scale of Fig.\ \ref{figarea}. Note that the estimated spin of LaCoO$_{2.5}$ (at T=913 K) is consistent with its known value.

\subsection{Testing the data with the different scenarios}

To determine which spin-transition scenario is in agreement with the data, the temperature behavior of magnetic susceptibility or anomalous volume expansion are usually analyzed in terms of thermally excited localized two or three-state models.\cite{saitoh1997,yamaguchi1997,radaelli2002,zobel2002} In terms of the three-state model, we can express the average spin on cobalt as the sum of the contribution of the (non-zero spin) excited states:

\begin{equation}
\bar{S}(T) = {\sum_{i=1}^2 S_i \nu_i (2S_i+1) e^{-\Delta_i/k_{B}T}} / Z,
\label{eqS}
\end{equation}

where $S_i$ is the spin of the excited state $i$, $\Delta_i$ is its energy with respect to the LS ground state, and $\nu_i$ is its orbital degeneracy, while $Z$ denotes the partition function. Thus, in the LS/IS/HS scenario $S_1=1,S_2=2$, while $\nu_i$ can have different values; for the two-state models the unnecessary term is omitted.

\begin{figure}  
\includegraphics[width=8.5cm]{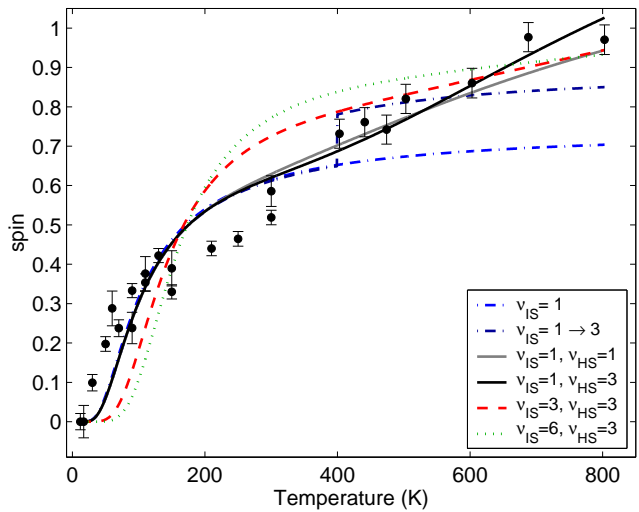}
\caption{\label{figfit} (color online) Fit of the temperature evaluation of the cobalt spin state with two- and three-state models described in the text. The legend describes the different orbital degeneracies assumed in the particular models.}
\end{figure}

Equation \ref{eqS} was fitted to the obtained spin data (presented in the top of Fig.\ \ref{figarea}) to determine the model and orbital degeneracies compatible with it. In Fig.\ \ref{figfit} we compare our experimental results on the spin of cobalt and fits to the various models. 
The simple two-state models do not give an appropriate description, especially the fits with the LS/HS scenario, and those of the LS/IS scenario with higher IS orbital degeneracy ($\nu_{\mathrm{IS}}>1$) were so poor that they are not included in the discussion. As even the remaining possibile two-state model, which involves a transition to the IS excited state with $\nu_{\mathrm{IS}}=1$, gives a poor fit at high temperatures (see the dashed line), a model with a change in the IS orbital degeneracy ($\nu_{\mathrm{IS}}=1\rightarrow3$) at high temperatures was also tested. Such ''switching on'' of the orbital degeneracy at high temperature was suggested by Zobel \textit{et al.}\ to attain a better agreement with susceptibility data.\cite{zobel2002} However, this model, which is shown in Fig.\ \ref{figfit} as a dashdot line with a jump at 400 K, does not bring a satisfactory improvement to the fit of the high temperature data.
Among the three-state-models, the ones with an orbitally nondegenerate ($\nu_{\mathrm{IS}}=1$) IS first excited state follow most reasonably the temperature evolution of the spin state (solid lines). The necessity of invoking the third state is evident from a comparison of these lines with that of the previously described two-state models. 
The fits with the other three-state models that involve higher orbital degeneracy increase too rapidly, so they give poor description for the entire temperature region (dotted and dashed lines). With these models, varying the orbital degeneracy of the HS state does not yield a relevant difference, and thus, for clarity, only those with ($\nu_{\mathrm{IS}}=3$), which give a slightly better fit, are included in the figure. In brief, in contrast to recent susceptibility measurements,\cite{zobel2002, baier2005} the data is best described with the three-state model.
The parameters of the fits are shown in Table \ref{tabfit}, the description of its lines follow the description of the lines in Fig.\ \ref{figfit} presented above.
In summary, three-state models (LS/IS/HS) with an orbitally nondegenerate ($\nu$=1) IS first excited state give the best description of the temperature evolution of the spin state.  The energy gaps obtained for the $\nu$=1 IS cases, which correspond to temperatures of 188-194 K, are in good agreement with the values of 180~K and 185~K, obtained by Zobel \textit{et al.}\ recently from magnetic susceptibility measurements and from anomalous thermal expansion, respectively,\cite{zobel2002} and with the 180~K obtained with $^{59}$Co NMR.\cite{kobayashi2000}

\begin{table}
\caption{\label{tabfit} Model orbital degeneracies and fit parameters for the spin on cobalt. The first line describes the LS/IS two-state model, the second is the LS/IS scenario with a change in the orbital degeneracy above 400K. The other lines belong to LS/IS/HS three-state models. }
\begin{ruledtabular}
\begin{tabular}{lllclc}
$\nu_{\textrm{IS}}$ & $\nu_{\textrm{HS}}$ & $\Delta_{\textrm{IS}}$ / meV & $\Delta_{\textrm{HS}}$  / meV & $\Delta_{\textrm{IS}}$ / K & $\Delta_{\textrm{HS}}$  / K \\ \hline
1 & - & 16.2 & -  & 188 & -  \\
1$\rightarrow$3 & - & 16.6$\rightarrow$32 & - & 193$\rightarrow$371 & -  \\
1 & 1 & 16.7 & 130 & 194 & 1504  \\
1 & 3 & 16.6 & 180 & 193 & 2085  \\
3 & 1 & 31.8 & 187 & 369 & 2164  \\
3 & 3 & 31.8 & 223 & 369 & 2592  \\
6 & 1 & 43.2 & 253 & 501 & 2938  \\
6 & 3 & 43.2 & 278 & 501 & 3223  \\
\end{tabular}
\end{ruledtabular}
\end{table}

It should be noted, however, that we assumed the energy gaps $\Delta_i$ to be independent of temperature. Albeit this conservative approach was applied in many recent works with success,\cite{saitoh1997,yamaguchi1997,zobel2002} modifications of the crystal-field splitting and hybridization strength with temperature are likely to affect the energy of the excited states. Hence, based on a description proposed long ago,\cite{bari1972} Asai \textit{et al.}\ applied corrections to the spin-state energies, which varied with the extent of the transformation,\cite{asai1998} while Baier \textit{et al.}\ suggested a straightforward way to extract the temperature dependence from the experimental data at the cost of restraining themselves to the two-state model. In the latter, the extracted $\Delta(T)$ exhibits a moderate variation at low temperatures, while above the insulator-metal transition ($T$\textgreater 500~K) it drops rapidly.\cite{baier2005} Applied to our data with the appropriate reformulation of Eq.\ \ref{eqS}, the results of this approach suggest different high-temperature behavior (namely, a more rapid decrease of $\Delta(T)$ above room temperature) than that obtained from the susceptibility data.\cite{baier2005} 
Although we do not favor analyzing the data with this approach due to its strong restrictions, we also refrain here from using the other description,\cite{asai1998} since temperature-dependent gaps would require introducing more fit parameters, and this would be challenged by the limitations of the data set presented here. Nevertheless, with the "conservative" approach a reasonable description of the experimental data was achieved, thus we expect that the variation of $\Delta_i$ should not be dramatic. Actually, many aspects of the temperature behavior of LaCoO$_3$ support this (especially below the insulator-metal transition), which include relatively small variations of the relevant bond lengths,\cite{radaelli2002} small excess entropy,\cite{stolen1997} and relatively small changes in the vibrational spectra\cite{rao2004}.  A more detailed study on LaCoO$_{3}$ and its partially substituted compounds will follow this work, which should permit us to address the nature of the temperature dependence of the higher spin state(s) using more refined models.

\section{Conclusions}

We have studied the temperature and pressure induced spin-state transitions of LaCoO$_3$ with high resolution x-ray spectroscopies. The XE spectra reflects continuous temperature variation of the spin localized on the cobalt. Similarly, the PFY XANES spectra shows the continuous redistribution of the $3d$ electrons between the $t_{2g}$ and $e_g$ levels of cobalt. We have developed a model-independent way to quantitatively follow the spin transition with X-ray emission spectroscopy through the integrated absolute difference spectra. Our high-pressure study confirmed that the pressure-induced transition reported by Vogt \textit{et al.}\cite{vogt2003} is indeed a transition to the LS state. 
The temperature behavior of the cobalt spin was found to be most compatible with a transition to an orbitally nondegenerate IS state with a gap of about 190~K. The higher transition requires invoking new parameters into the description; our data was best described with a partial transition to a HS state. Future detailed experiments utilizing resonant and non-resonant X-ray emission spectroscopy should shed more light on the variations of the electronic and spin state.

\begin{acknowledgments}
The authors thank Prof.~L.~H.~Tjeng, Prof.~M.~Abd-Elmeguid and Dr.~G.~Monaco for stimulating discussions, Drs. T. Lorenz and C. Zobel for providing the LaCoO$_3$ samples, Dr. S. Huotari for his careful reading of the manuscript, Dr.~R.~Verbeni for the preparation of the analyzer crystal, and C.~Henriquet for technical assistance.
\end{acknowledgments}


\end{document}